\documentclass[a4paper]{jpconf}
\usepackage{graphicx}

\newcommand {\I}{\mathbf{I}}
\newcommand {\D}{\mathbf{D}}
\newcommand {\T}{\mathbf{\Theta}}

\newcommand {\G}{\mathbf{G}}
\newcommand {\V}{\mathbf{V}}

\newcommand {\bS}{\mathbf{S}}
\newcommand {\bP}{\mathbf{P}}

\newcommand {\bk}{\mathbf{k}}
\newcommand {\re}{\mathbf{r}}

\newcommand {\R}{\mathbf{R}}
\newcommand {\M}{\mathbf{M}}

\newcommand {\sym}{\mathbf{Sym}}

\newcommand {\bb}{\mathbf{b}}
\newcommand {\sgn}{\mathbf{s}}
\newcommand {\Sgs}{\mathbf{S}}
\newcommand {\Sgm}{\mathbf{S_m}}

\begin{document}
%%%%%%%%%%%%%%%%%%%%%%%%%%%%%%%%%%%%%%%%%%%%%%%%%%%%%%%%%%%%%%%%%%%%%%%%
%%%%%%%%%%%%%%%%%%%%%%%%%%%%%%%%%%%%%%%%%%%%%%%%%%%%%%%%%%%%%%%%%%%%%%%%

%%%%%%%%%%%%%%%%%%%%%%%%%%%%%%%%%%%%%%%%%%%%%%%%%%%%%%%%%%%%%%%%%%%%%%%%
\title{Simplification of tensor expressions in computer algebra}
%%%%%%%%%%%%%%%%%%%%%%%%%%%%%%%%%%%%%%%%%%%%%%%%%%%%%%%%%%%%%%%%%%%%%%%%

\author{A.Kryukov and G.Shpiz}

\address{Skobeltsyn Institute of Nuclear Physics Lomonosov Moscow State University, \\ 119992 Moscow, Russia}

\ead{kryukov@theory.sinp.msu.ru}

%%%%%%%%%%%%%%%%%%%%%%%%%%%%%%%%%%%%%%%%%%%%%%%%%%%%%%%%%%%%%%%%%%%%%%%%
\begin{abstract}
%%%%%%%%%%%%%%%%%%%%%%%%%%%%%%%%%%%%%%%%%%%%%%%%%%%%%%%%%%%%%%%%%%%%%%%%

Computer algebra is widely used in various fields of mathematics, physics and other sciences.
The simplification of tensor expressions is an important special case of computer algebra. 
In this paper, we consider the reduction of tensor polynomials to canonical form, taking into account the properties of symmetry under permutations of indices, the symmetries associated with the renaming of summation indices, and also linear relations between tensors of a general form. 
We give a definition of the canonical representation for polynomial (multiplicative) expressions of variables with abstract indices, which is the result of averaging of the original expression by the action of some finite group (the signature stabilizer). 
In practice, the proposed algorithms demonstrate high efficiency for expressions made of Riemann curvature tensors.

\end{abstract}

%%%%%%%%%%%%%%%%%%%%%%%%%%%%%%%%%%%%%%%%%%%%%%%%%%%%%%%%%%%%%%%%%%%%%%%%
%%%%%%%%%%%%%%%%%%%%%%%%%%%%%%%%%%%%%%%%%%%%%%%%%%%%%%%%%%%%%%%%%%%%%%%%
\section{Introduction}
\label{intro}
%%%%%%%%%%%%%%%%%%%%%%%%%%%%%%%%%%%%%%%%%%%%%%%%%%%%%%%%%%%%%%%%%%%%%%%%

Tensor calculations are widely used in theoretical physics, solid state physics, 
mechanics and many other fields.

There are three main approaches to the simplification of tensor expressions which are used in computer algebra: (i) component calculations, (ii) manipulations with tensors with abstract indices and (iii) abstract tensor calculations.
All of them have some advantages and disadvantages. 
The easiest approach to the tensor simplification is  the component calculations. 
With this approach we have to choose some basis. 
All further calculations in this basis are the calculations of components of available tensors in a specific representation. 
Thus this approach reduces the task to calculations of scalar values the number of which is equal to the number of tensor components. 
The disadvantage of this approach is the huge number of tensor components that we must calculate.
Also we cannot use special tensor properties, such as symmetry, automatically.

As examples of such calculations, we can specify the programs xAct~\cite{Mar08}, DifferentialGeometry~\cite{And12}, GRTensor~\cite{GRT2}, Atlas~2~\cite{Atl2},CTENSOR~\cite{Tot05}.

The last type of tensor simplification is purely abstract calculus. A good example of this approach is exterior algebra. 
We will not discuss it here.
Examples of such programs are xTerior~\cite{xTer} in the system xAct~\cite{Mar08}, or ITENSOR~\cite{Tot05} in the CAS MAXIMA~\cite{MAX}.

In this paper we consider the second approach, in which a tensor is considered as an abstract object with indices, possessing various symmetry properties.
We briefly discuss some computer algebra algorithms for simplification of tensor polynomials. 
Also we present a sketch of an algorithm based on stabilizers of permutation subgroups and on isomorphism of graphs corresponding to tensor monomials.

In the paper we will use the Einstein notations implying that two identical indices in a monomial means summation over all their values:
$$
T(i,i) \equiv \sum_{i=1}^{i=n}T(i,i)
$$

For simplicity, we will not distinguish upper and lower indices.
Also we will not be interested in the transformation properties of tensors under coordinate transformations.

The structure of the paper is as follows. Section \ref{dcoset} briefly describes the use of the coset approach for tensor simplification.
Section \ref{galg} is devoted to the group algebra approach. 
In Section \ref{Simpl}, we give necessary definitions and statements.
In Section \ref{canon}, we describe the approach based on stabilizers of groups and isomorphism of graphs. We also give a sketch of the algorithm based on the ideas above.
In conclusion, we summarize the results obtained and present plans for further development.

%%%%%%%%%%%%%%%%%%%%%%%%%%%%%%%%%%%%%%%%%%%%%%%%%%%%%%%%%%%%%%%%%%%%%%%%%%
\section{Double coset}
\label{dcoset}
%%%%%%%%%%%%%%%%%%%%%%%%%%%%%%%%%%%%%%%%%%%%%%%%%%%%%%%%%%%%%%%%%%%%%%%%%

We will consider a tensor as an abstract object with indices $T(i,j,k,\ldots)$, which has certain symmetry properties with respect to the permutation of the indices.

For example, let us consider the Riemann tensor $R(i,j,k,l)$. 
The Riemann tensor has the following symmetry properties:
\begin{eqnarray}
R(i,j,k,l) = -R(j,i,k,l),  \label{Ri1} \\
R(i,j,k,l) = R(k,l,i,j).   \label{Ri2}
\end{eqnarray}

To give a general example, we can say that all symmetries of this kind form a subgroup of the permutation group of indices:
\[
S_N \subset P_N, 
\]
where $N$ is the order of the permutation group.

Another type of symmetry arising in tensor expressions is the symmetry associated with summation of indices. 
For example, consider the scalar curvature $R$:
\begin{displaymath}
R = R(i,i), \;\;where\;Ricci\;tensor\;\; R(i,j) = R(k,i,k,j).
\end{displaymath}

The scalar curvature $R$ has an additional symmetry associated with the renaming of summation indices
\begin{equation}
R = R(k,i,k,i) = R(i,k,i,k) = R(i_1,i_2)=R(i_2,i_1), \label{Bi1}
\end{equation}
where $i_1$ and $i_2$ are summation indices which were temporally marked by subindices $1$ and $2$ for illustration.

So summation indices generate another symmetry group which is a subgroup of the permutation group
\[
D_N \subset P_N
\]
and it acts from the right.

The problem of simplification of a tensor monomial which has both types of symmetry is reduced to the finding the double coset:
$$
S_N\backslash T \slash D_N.
$$

This approach was developed in \cite {Rod89} and is widely used in most tensor simplification algorithms (see, for example, \cite {But91, Man04}).

The problem is how to deal with so-called multi-term identities. For example, the Riemann tensor satisfies the following Bianchi identity:
\[
R(i,j,k,l)+R(i,l,j,k)+R(i,k,l,j)=0.
\]

Such identities destroy the group structure and must be treated by some other method.

%%%%%%%%%%%%%%%%%%%%%%%%%%%%%%%%%%%%%%%%%%%%%%%%%%%%%%%%%%%%%%%%%%%%%%%%
\section{Group algebra}
\label{galg}
%%%%%%%%%%%%%%%%%%%%%%%%%%%%%%%%%%%%%%%%%%%%%%%%%%%%%%%%%%%%%%%%%%%%%%%%

Another approach to the simplification of tensor expressions was proposed by V.~Ilyin and A.~Kryukov \cite{Ily96}. 
The package ATENSOR implements this idea in CAS REDUCE\cite{RED}.
In this approach all relations between tensor monomials both following from symmetries and multi-term identities were considered as vectors in some linear vector space $V$ with dimension $N!$. 

All the identities generate a hyperplane $H \subset V$.
The hyperplane $H$ defines a set of tensor expressions equivalent to zero.
Thus, the problem of simplifying a tensor expression is reduced to finding the (orthogonal) hyperplane $K$ that is supplementary to $H$.

Let us illustrate this point on the simplest example of symmetric tensor of the 2-nd rank $\delta(i,j)=\delta(j,i)$. 
We have two basis vectors: $e_1=\delta(i,j)$ and $e_2=\delta(j,i)$. Let $t$ denote the vector $t=\alpha e_1 + \beta e_2$. 
%(see Fig.~\ref{fig1}). 
The canonical representation of $t_k$ is a projection of $t$ on $K$
\[
t=\alpha e_1 + \beta e_2= \frac{1}{2}(\alpha-\beta)(e_1-e_2)+\frac{1}{2}(\alpha+\beta)(e_1+e_2)
\]
The first monomial is equal to zero. 
Thus the canonical form of $t$ is the following: $t=\frac{1}{2}(\alpha+\beta)(e_1+e_2) \in K$.
%\begin{figure}[h]
%\begin{center}
%\includegraphics[width=0.5\textwidth]{fig1.png}
%\end{center}
%\caption{Projection of $t$ on canonical element hyperspace $K$}
%\label{fig1}       % Give a unique label
%\end{figure}

The disadvantage of the group algebra approach is the huge dimension of the vector space equal to $N!$ where $N$ is the dimension of the full permutation group of tensor expressions. 
For example, a product of two Riemann tensors $R(i,j,k,l)\cdot R(k,l,m,n)$ has dimension of space equal to $8!=40320$.

%%%%%%%%%%%%%%%%%%%%%%%%%%%%%%%%%%%%%%%%%%%%%%%%%%%%%%%
\section{Definitions and statements}
\label{Simpl}
%%%%%%%%%%%%%%%%%%%%%%%%%%%%%%%%%%%%%%%%%%%%%%%%%%%%%%%

Let us fix two linearly ordered sets: the set of types $\T$ and the set of indices $\I$.
In the set $\I$, we fix a subset $\D \subset \I$ of the summation indices.

We will call the \textbf{signature} the pair $(t,i)$, where $t\in \T$, and $i=(i_1,\dots,i_n)$ is non-decreasing sequence of indices.

\textbf{Elementary symbol} of order $n$ is an expression $e$ of the form $e = t(i)$, where $t \in \T$,$\i = (i_1, \dots, i_n)$,$i_k \in \I$.
 
\textbf{Signature of symbol} $e=t(i_1,\dots,i_n)$ will be the pair $(t,i')$, where $t\in \T$, and $i'=(i'_1,\dots,i'_n)$ ordered by ascending sequence of the indices $i=(i_1,\dots,i_n)$.

Let $\Sgs$ denote the set of all symbol signatures.
  
The set of sequences of indices is linearly ordered with respect to the lexicographic order.
The set of signatures and the set of elementary symbols are also linearly ordered with respect to the lexicographic order.
 
The vector space of formal linear combinations of elementary symbols will be denoted by $\sym$, and its elements will be called indexed values.

Let $s=(t,\mathbf{i})\in \Sgs$ be a signature.

We denote by $\sym(s) \subset \sym$ the vector space generated by all elementary symbols with the signature $s$, that is, from characters of type $t$ and a sequence of indices obtained from the sequence $i = (i_1, \ dots, i_n)$ by some permutation of elements.

The group $G$ of permutations of the set of indices $\I$ naturally acts on the set of all elementary symbols and, accordingly, in the space $\sym$. In the following, we will call it the index renaming group. If $g \in G$, then $\sym (ge) = g(\sym(e))$.

Obviously, the index renaming group $G$ translates symbols with the same signature into symbols with the same signature, and its natural action on the set of signatures $\bS$ is defined.

Let $D \subset \I$ be some subset of indices. We denote by $G(D)$ a subgroup of the index renaming group that converts the set $D$ into itself leaving indices from the complement of $D$ in place.

The group of renaming of summation indices will be denoted $G_d$.

We assume that all relations between elementary symbols are reduced to some linear relations between symbols with the same signature, and the set of relations is invariant with respect to the index renaming group.

Let $\R(s)$ denote the space of all (linear) relations in the space $\sym(s)$.
We will require that the correspondence $s\to \R(s)$ be invariant with respect to the renaming of the indices.

Let $\R$ denote the subspace in $\S$ generated by all $\R(s)$ in the space $\sym$. 
By construction, $\R$ is invariant with respect to renaming of indices and $\R(s) = \sym(s) \cap{\R}$.

In the space $\sym(s)$ there is a single sequence $\bb (s) = e_1, \dots, e_k$ of elementary symbols forming the basis in the factor space $\V(s) = \sym(s) / \R(s)$, which is minimal among such sequences in the sense of lexicographical order.

The space generated by the basis $\bb(s)$ is naturally identified by the quotient $\sym(s) / \R(s)$ and will be denoted $\V(s)$.

We will call $\bb(s)$ the \textbf{canonical basis}, and elements of the canonical basis will be called \textbf{canonical symbols}.

\textbf{Example}. Let $s = (Ri,(1,2,3,4))$ be the signature of the Riemann tensor.
Then the canonical basis in $\V(s)$ consists of two symbols: $Ri(1,2,3,4)$ and $Ri(1,3,2,4)$.

Since the space of relations $\R(s)$ is invariant with respect to the group $\G(s)$ of renaming the indices entering $s$, this group acts in $\V(s) = \sym(s) / \R(s)$.
Note that with this action, the elementary symbol turns, generally speaking, into a linear combination of canonical symbols.

\textbf{Monomial} is the formal commutative product of elementary symbols.
We will require that the summation indices and only they be in the set $\D$.

Let $\M$ denote the set of all monomials, and $\bP$ be a vector space with basis $\M$.
Elements of the space $\bP$ are naturally called \textbf{polynomials}.

\textbf{The Signature} $\sgn(m)$ of a monomial $m = e_1 \cdot ... \cdot e_k$ is the sequence of signatures of its factors, ordered in ascending order.
 
The set of all possible signatures of monomials will be denoted by $\Sgm$.
 
Let $S$ be some set of monomial signatures.
Let $\M(S)$ denote the set of all monomials whose signatures lie in $S$ and 
 $\bP(S)$ be a vector space with basis $\M(S)$.
 
Let $m = e_1 \cdot ... \cdot e_k$ be a monomial.
Denote by $\bS(m)$ the set of signatures of all monomials obtained from $\sgn(m)$ by renaming the summation indices, that is, the orbit of $\sgn(m)$ with respect to the action of the group $\G_d$.

There are three types of linear relations between monomials.
\begin {enumerate}
\item Relations generated by relationships within one of the factors, with the other factors fixed.
\item Equality of monomials obtained by permuting the factors
\item Equality of monomials obtained by renaming summation indices (that is, by the action of the group$\G_d$).
\end {enumerate}

Denote by $\bP(m)$ the vector space generated by all monomials whose signatures lie in $\bS(m)$, and $\R(m)\subset\bP(m)$ is the subspace generated by relations of the types listed.
The subspace generated by relations of type (i) and (ii) will be denoted by $\R_0(m)$.

The spaces $\bP(m)$,$\R(m)$ and $\R_0(m)$ are invariant with respect to the action of the group $\G_d$.
In particular, this group acts on the factor spaces $\bP(m)/ \R_0(m)$ and $\bP(m) / \R(m)$.

We call a monomial \textbf{reduced} if all its factors are canonical symbols and the sequence of factors is ordered in ascending order.

The space generated by all reduced monomials in $\bP(m)$ will be denoted by $\bP_r(m)$.

Any monomial in the space $\bP(m)$ can be represented up to a relation in $\R_0(m)$ as a linear combination of reduced monomials.
The space $\bP_r(m)$ is naturally identified with the factor-spaces $\bP(m) / \R_0(m)$.

We denote by $\rho: \bP(m) \to \bP_r(m)$ the natural projector onto the factor space
 $\bP(m) / \R_0(m)$.
 
Note that the image $\rho(m)$ of the monomial $m$ is generally a polynomial.

%%%%%%%%%%%%%%%%%%%%%%%%%%%%%%%%%%%%%%%%%%%%%%%%%%%
\section{Reducing monomials to the canonical form}
\label{canon}
%%%%%%%%%%%%%%%%%%%%%%%%%%%%%%%%%%%%%%%%%%%%%%%%%%%

\textbf{The canonical representation} is a linear operator $\kappa: \bP \to \bP$ 
such that $x = \kappa(x)\; mod(\R)$ and $\kappa(x) = \kappa(y)$ 
if and only if $x = y\; mod(\R)$ for $x,y \in \bP$.

We define the operator $\alpha: \bP \to \bP$, assuming for arbitrary $p \in \bP$ that
\begin{eqnarray}
\alpha(p) = \frac{1}{N} \sum_{g \in \G_d} g(p) \label{a_op}
\end{eqnarray}
where $N$ is the number of elements in the group $\ G_d$.

The operator $\alpha \rho = \rho \alpha$ is the canonical representation.

Unfortunately, this canonical representation is too cumbersome for real computations.
Therefore, we modify its structure.

The set $\Sgm$ of the signatures of all monomials is a subset of the set of finite sequences of signatures of elementary symbols linearly ordered with respect to the lexicographic order.
Accordingly, the set $\Sgm$ is linearly ordered with respect to the inherited order.

The group $\G_d$ of renaming of summation indices acts naturally on the set of monomials and on the set of their signatures.

The minimal element in the orbit of the signature $\sgn(m)$ of the monomial $m$ with respect to the action of the group $\G_d$ will be denoted by  $\sgn_{min}(m)$.

The signature of the monomial $m$ will be called \textbf{extremal} one if it cannot be reduced by any renaming of the summation indices, that is, it coincides with $\sgn_{min}(m)$.

A monomial will be called \textbf{extremal} if its signature is the extremal one.

The space generated by all extremal reduced monomials in $\bP$ will be denoted by $\bP_{min}$.
It coincides with the subspace in $\bP_{min}(m)$ generated by all monomials with signature $\sgn_{min}(m)$.

Any monomial $m \in \bP_{min}$ can be represented up to a relation in $\R$ as a linear combination of extremal reduced monomials with the signature $\sgn_{min}(m)$.

Let the set of monomials $M \in \bP_{min}$ contain one monomial from $\bP_{min}$ for each signature.
The space $\bP_{min}$ is a direct sum
$$
\bigoplus_{m \in M} \bP_{min}(m)
$$
and
$$
\R \cap \bP_{min} = \bigoplus_{m \in M} \R \cap \bP_{min}(m).
$$

For the monomial $m \in \bP$ we set $\re(m) = \{x \in \bP_{min}(m): x = m\;mod(\R) \}$. Thus, $\re(m)$ is the set of all polynomials from $\bP_{min}(m)$ that coincide with $m$ modulo relations from $\R$.

Denote by $\bk(m)$ the arithmetic mean of the elements from the set $\re(m)$.

Obviously, the function $\bk(m)$ extends to a linear operator on the space $\bP$, which we will denote by the same symbol $\bk: \bP \to \bP_{min}$.

The operator $\bk$ is a canonical representation.

Note that the dimension of the space $\bP_{min}$ is significantly less than the dimension of the space $\bP$, and the signature stabilizer $\sgn_{min}(m)$ in the group $\G_d$ is much smaller than the group itself.

%%%%%%%%%%%%%%%%%%%%%%%%%%%%%%%%%%%%%%%%%%%%%%%%%%%%%%%%%%%%%%%%%%%
\subsection{The algorithm}
\label{gIso}
%%%%%%%%%%%%%%%%%%%%%%%%%%%%%%%%%%%%%%%%%%%%%%%%%%%%%%%%%%%%%%%%%%%

We assume that at each step of the simplifications: (1) all elementary tensors are reduced to the canonical form;
(2) there are no contraction of indices into an elementary tensor. We also use lexicographically order in the tensors.

We will map each monomial to the colored graph.
The vertices of such graphs correspond to elementary symbols that are factors of monomials, the colors of the vertices being the signatures of the corresponding factors from which the summation indices are excluded, and the edges correspond to the summation indices.

If the monomials are mapped into two non-isomorphic graphs, then the monomials lie in non-intersecting spaces modulo the relations (independent). 

\subsection{The algorithm(sketch)}

\begin{enumerate}
\item Transform each tensor term to the canonical form with respect to its symmetry properties. 
Every step should be finished with the canonization of tensor itself and ordering the tensors in the monomials (so called “pre-canonical form”).
\item Apply the allowed transformation of summation indices to the monomial.
\item Calculate the average over all the monomials which have isomorphic graphs with the initial one.
\item The obtained polynomial is invariant with respect to the transformations from the equivalence class of the initial
monomial.
\end{enumerate}

Using the obtained invariants we can get the canonical form of the monomials.

%%%%%%%%%%%%%%%%%%%%%%%%%%%%%%%%%%%%%%%%%%%%%%%%%%%%%%%%%%%%%%%%%%%%%
\section*{Conclusion}
%%%%%%%%%%%%%%%%%%%%%%%%%%%%%%%%%%%%%%%%%%%%%%%%%%%%%%%%%%%%%%%%%%%%%%

The proposed algorithm significantly reduce the amount of computations in the case of large groups.

We have implemented a prototype of the program on Python. 
The program is working rather good for tensor expressions in practical cases where monomials
contain about 10--15 factors and each factor contains up to 8--10 indices. 
The next step is a further program code refinement and optimization.

Our nearest plan is to finish the development of the program and to carry out a detailed investigation of the algorithm as well as to compare it with other algorithms for tensor simplifications available on the market.

\ack{
The authors are grateful to V.~Ilyin and A.Demichev for his interest in the work and useful discussions. 
We are also grateful to J.~Dubenskaya for help of preparing of the paper.
}

%%%%%%%%%%%%%%%%%%%%%%%%%%%%%%%%%%%%%%%%%%%
%%%%%%%%%%%%%%%%%%%%%%%%%%%%%%%%%%%%%%%%%%%
\section*{References}
%%%%%%%%%%%%%%%%%%%%%%%%%%%%%%%%%%%%%%%%%%%
%%%%%%%%%%%%%%%%%%%%%%%%%%%%%%%%%%%%%%%%%%%

%%%%%%%%%%%%%%%%%%%%%%%%%%%%%%%%%%%%%%%%%%%%%%%%%%%%%%%%%%%%%%%%%%%%%%%%

%%%%%%%%%%%%%%%%%%%%%%%%%%%%%%%%%%%%%%%%%%%%%%%%%%%%%%%%%%%%%%%%%%%%%%%%
\end{document}